# The fractal spatial distribution of stars in open clusters and stellar associations

Néstor Sánchez, Emilio J. Alfaro

*Instituto de Astrofísica de Andalucía, CSIC, Apdo. 3004, E-18080, Granada, Spain*

**Abstract:** The Interstellar Medium has a fractal structure, in the sense that gas and dust distribute in a hierarchical and self-similar manner. Stars in new-born cluster probably follow the same fractal patterns of their parent molecular clouds. Moreover, it seems that older clusters tend to distribute their stars with radial density profiles. Thus, it is expected that clusters form with an initial fractal distribution of stars that eventually evolves toward centrally concentrated distributions. Is this really the case? This simple picture on to the origin and early evolution of star clusters and associations is very far from being clearly understood. There can be both young clusters exhibiting radial patterns and evolved clusters showing fractal structure. Additionally, the fractal structure of some open clusters is very different from that of the Interstellar Medium in the Milky Way. Here we summarize and discuss observational and numerical evidences concerning this subject.

## 1. INTRODUCTION

Nowadays it is known that most stars form from the gravitational collapse of gas and dust in large complexes of giant molecular clouds (GMCs). The overall picture is widely accepted among astronomers. Gas inside GMCs is protected from interstellar ultraviolet radiation field and it can cool and collapse into a number of protostellar objects. The mass and temperature of each protostar increase due to accretion of gas from the surrounding medium until hydrogen begins to fuse in the core and a star is born. Stars do not form individually but in groups (Lada and Lada, 2003), and the stellar winds and jets that emerge from these young stars shock and blow away the surrounding gas leaving behind the newly formed star cluster or group of clusters.

This is, however, only a general description. There are still many fundamental facts that are unknown. Recent reviews about what is known (or unknown) on the star formation process are those by Larson (2003), Mac Low and Klessen (2004) and McKee and Ostriker (2007). One important example of this is the distribution of the mass of the new-born stars, i.e. the initial mass function (IMF). The way the gas matter distributes into stars to produce a power-law distribution is far to be known and cannot be derives from this global scenario. Despite all the theoretical and observational knowledge regarding this subject, it is unclear whether the IMF is a universal function or whether, on the contrary, is sensitive in some way to initial and/or environmental conditions (Bastian et al., 2010). In fact, it is not known whether the shape of the IMF comes directly from the internal mass distribution of the GMC or whether it is determined or modified from other physical mechanisms during the star formation process (Bonnel et al., 2007). In other words, at this time it is not known what physical processes are responsible for the observed stellar masses. The situation is even worse for stars more massive than ~ 8 $M_\odot$ where there is much debate about the mechanisms dominating the formation, such as monolithic collapse, competitive accretion or stellar mergers (Zinnecker and Yorke, 2007).

The main reason for this lack of knowledge is the large complexity of the star formation process. The initial conditions, that is, the properties of the cold and dark clouds that eventually form stars are poorly known (a recent review is in Bergin and Tafalla, 2007). The structure of the density and velocity fields seems to be determined mainly, but probably not exclusively, by turbulent motions occurring within the clouds although the main physical sources of turbulence are not yet fully understood (Elmegreen and Scalo, 2004). Turbulence plays a dual role. On the one hand, it tends to dissipate the gas inhibiting or delaying the gravitational collapse. On the other hand, it can originate high-density, shocked regions that can undergo gravitational collapse and form stars. Besides turbulence, self-gravitation has to act against tidal forces at large spatial scales and against magnetic forces and/or thermal pressure at smaller scales in order to form stars. The physical mechanisms that seem to be basic to form stars (turbulence, magnetic fields, and self-gravity) are themselves highly nonlinear. Stellar feedback is also an important factor (Larson, 2003; McKee and Ostriker, 2007). The injection of energy, mass and momentum into the medium by the stars may disperse the gas cloud preventing the formation of other stars. But stellar feedback may also stimulate the formation of stars compressing the surrounding gas and triggering the gravitational collapse. Even if very few physical processes are considered, the interrelationship among the different processes and the interaction among different regions of the cloud and different spatial scales are such that the star formation process is likely very sensitive to small variations on the initial or environmental conditions (Sánchez and Parravano, 1999).

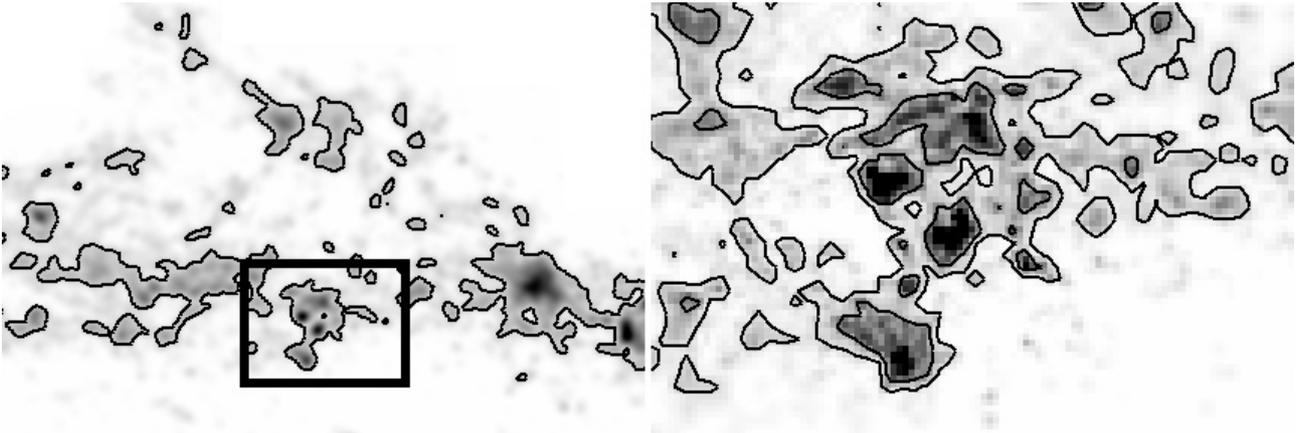

Figure 1: Velocity-integrated CO image of part of the Milky Way. The gray scale runs linearly from 0 (white) to 40 K km s$^{-1}$ (black). Left, the region with Galactic coordinates 83º 00' ≤ $l$ ≤ 136º 30' and -08º 30' ≤ $b$ ≤ +25º 45' is fragmented into different regions. Right, a detailed view of a single fragment (square area in the left image) shows smaller fragments contained in it.

There are also observational limitations. Stars are born embedded within GMCs and observations of the formation process and earliest stages are heavily affected by extinction. Important advances have been made studying young embedded clusters, mainly in the radio and infrared ranges (Lada and Lada, 2003). In any case, the time scales involved in star formation are such that it is only possible to see snapshots at different stages in different regions with different properties and from that try to infer the physics behind the whole process. Knowing the structure of GMCs gives information about the initial conditions prior to the beginning of the process. The study of young star clusters yields useful information about the stage just after stars are born. The analysis and comparison of these two approximations may yield important clues regarding the star formation process itself. Here we use this approach to review and discuss the initial distribution of stars in open clusters and stellar associations.

The observed distribution of stars in an open cluster or association, i.e. the so-called internal structure, is the result of some initial distribution and its subsequent dynamical evolution. Star formation occurs mainly along the densest regions of the GMCs. Thus, in the simplest picture, we expect the initial distribution of stars to be given by or related to the distribution of density peaks in their parent GMC. Let us start reviewing what we know about the internal structure of molecular clouds.

## 2. FRACTAL STRUCTURE OF THE INTERSTELLAR MEDIUM

Gas and dust in the interstellar medium (ISM) are not uniformly distributed in the Galaxy. They are organized into irregular structures in a hierarchical and approximately self-similar manner. This means that each structure is composed of smaller and very similar structures which are composed of even smaller structures and so on. As an example, Figure 1 shows an integrated CO emission map of part of the Milky Way, from the data of the Whole-Galaxy CO Survey by Dame et al. (2001) available at http://www.cfa.harvard.edu/mmw/MilkyWayinMolClouds.html. The overall structure is quite similar at different spatial scales. This self-similarity is a typical property of geometric objects called "fractals" (Mandelbrot, 1983). In a fractal, each small part is, at least approximately, a reduced copy of the whole object. For this reason it is often said that interstellar clouds exhibit fractal properties. Obviously, the ISM does not behave as a perfect fractal in the mathematical sense, for instance it has not substructures at arbitrarily small scales. However, it can be said that *in a certain range of spatial scales* the ISM can be well *described* as a fractal structure.

It is not a trivial task to characterize in a quantitative and therefore objective way the complexity of structures as those seen in Figure 1. One common strategy is to measure mass and size of certain types of pre-defined objects in the hierarchy. Dense "cores", with sizes of the order of ~ 0.1 pc and masses of ~ 1 M$_{\odot}$ are in the lower levels of the hierarchy. These cores are known as pre-stellar objects and will eventually collapse forming individual stars or multiple systems. At a higher level are "clumps" with sizes of ~ 1 pc and masses in the range ~ 10-100 M$_{\odot}$. Molecular "clouds" with sizes of ~ 10 pc contain several clumps and cores with masses in the range ~ $10^3$-$10^4$ M$_{\odot}$ (Bergin and Tafalla, 2007). These structures usually exist inside giant molecular clouds of ~ 100 pc with ~ $10^6$ M$_{\odot}$, which can be grouped forming large complexes or fragments of spiral arms with sizes of the order of kpc and masses up to ~ $10^7$ M$_{\odot}$ (Efremov, 2010). These are typical values for objects for which, actually, there exist no precise definitions because in a rigorously hierarchical scenario there are no characteristic spatial scales that can be used to define any particular structure.

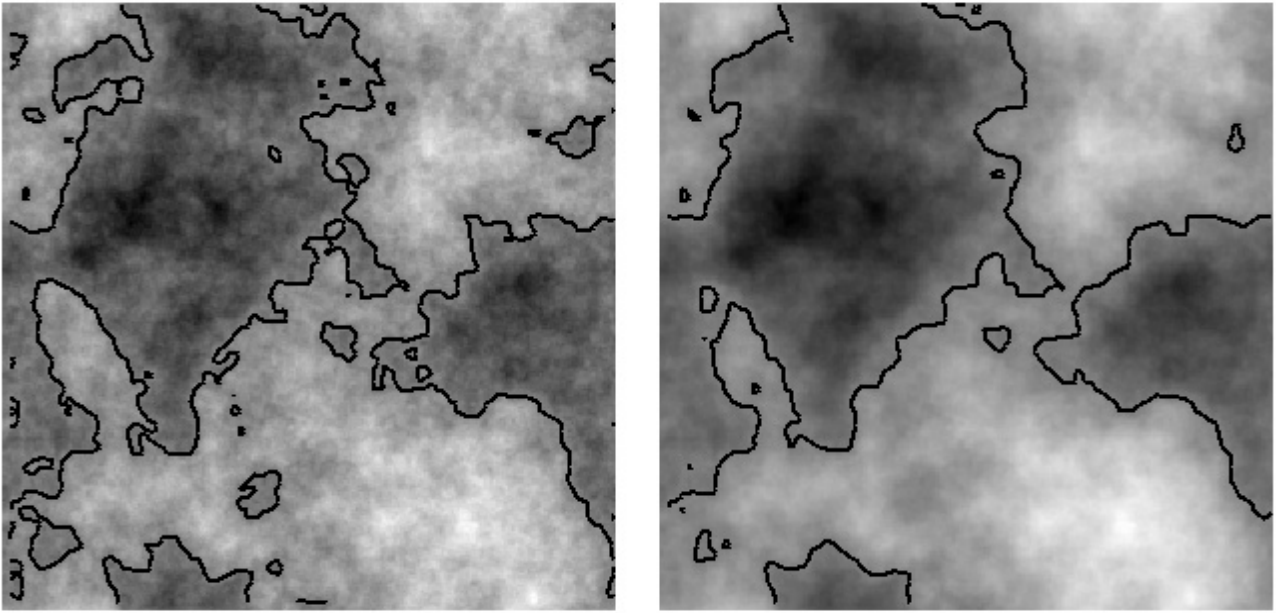

Figure 2: Two maps of simulated two-dimensional fractal clouds. The clouds have exactly the same intrinsic gas distribution but different fractal dimensions: $D_{per} = 1.5$ (left) and $D_{per} = 1.2$ (right). Isocontours in both cases are drawn at 50% of the maximum intensity level. Perimeters on the left-hand side map are more convolved than those on the right-hand side map.

Many other tools are now widely used to describe the complexity of a region in the ISM as a whole system without defining things such as "cloud" (a comprehensive review can be found in Elmegreen and Scalo, 2004). Fractal analysis is particularly appropriate for dealing with hierarchical and self-similar (i.e. fractal) systems. The fractal dimension $D_f$ is a quantity that gives the degree of irregularity of an object. In the three-dimensional space, a cloud of gas and dust distributed in a smooth (homogeneous) manner will have a fractal dimension $D_f = 3$ (or $D_f = 2$ in a two-dimensional space, for instance an observed map of the cloud). In a fractal distribution there are dense fragments separated by less dense regions which in turn are formed by smaller fragments so that the matter exhibits an irregular and clumpy (heterogeneous) structure. In this last case the fractal dimension will be $D_f < 3$ (or $D_f < 2$ in two dimensions). The more irregular or far from homogeneity is the cloud, the smaller fractal dimension values. Thus, beyond the details on the characteristics of cloud fragments, the fractal dimension gives us the average degree of clumpiness of a given region.

What is the fractal dimension of interstellar clouds in our Galaxy? There are some problems with measuring it. We cannot see their three-dimensional (3D) shape because interstellar clouds are necessarily recorded as two-dimensional (2D) images projected onto the sky. Moreover, we cannot see the photons coming from the central part or from the back side of the cloud if the gas is very opaque to its own radiation. In the opposite case of transparent clouds, all the photons are observed but the information is mixed and distorted. All these effects will complicate the characterization of the cloud structure. A commonly used method consists of calculating the fractal dimension of the projected boundaries through the so-called perimeter-area-based dimension ($D_{per}$). If the isocontours exhibit a power-law relation between the perimeter $P$ and the enclosed area $A$ of the form $P \sim A^{D_{per}/2}$ then $D_{per}$ will be the fractal dimension of the isocontours (Mandelbrot, 1983). $D_{per}$ is simply the fractal dimension that characterizes the manner in which the projected boundaries fill space. Objects with smoothly varying contours (e.g., circles) have $D_{per} = 1$ ($P \sim A^{1/2}$) whereas extremely convolved plane-filling contours have $D_{per} = 2$ ($P \sim A$). This can be seen in the two fractal clouds simulated in Figure 2. In an interstellar cloud with the same underlying fractality through the whole structure and over a wide range of spatial scales (called a *monofractal*), the dimension of the contours will be the same for both small high-density cores and large low-density regions. Thus, an advantage of the perimeter dimension is that it is practically independent on cloud opacity (Sánchez et al., 2007a).

Lovejoy (1982) was the first to apply this procedure to Earth's clouds obtaining $D_{per} \simeq 1.35$. The first application to interstellar clouds was by Beech (1987), who obtained $D_{per} \simeq 1.4$ for a set of 24 selected dark clouds. Since then this method has been applied to different regions in the Galaxy observed at different wavelengths. The results show that the ISM behaves as a fractal in a wide range of spatial scales, namely from at least $\sim 0.01$ up to $\sim 1000$ pc. There is, however, a wide variation in the estimated fractal dimensions. A summary of results that can be accessed via NASA's ADS service is shown in Table 1 and is also presented graphically in Figure 3.

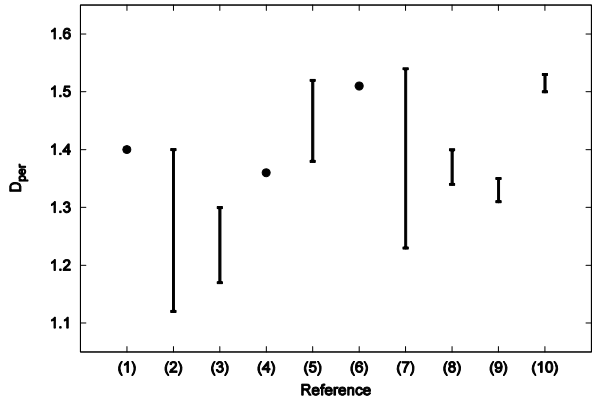

Figure 3: Summary of perimeter dimension values for several interstellar clouds in the Galaxy taken from literature. The horizontal axis gives the bibliographical reference (Table 1) and the vertical axis is the value (point) or range of values (bar) of $D_{per}$ for the different regions or maps.

Table 1: Summary of perimeter dimensions in molecular clouds in the Galaxy

| Ref. | $D_{per}$ | Region/Map |
|------|-----------|------------|
| (1) | 1.40 | Extinction maps of dark clouds |
| (2) | 1.12-1.40 | Dust emission maps of cirrus clouds |
| (3) | 1.17-1.30 | Infrared intensity and column density maps of several molecular clouds |
| (4) | 1.36 | Molecular emission maps (Taurus complex) |
| (5) | 1.38-1.52 | Visual extinction maps (Chamaeleon complex) |
| (6) | 1.51 | Molecular emission map (Taurus complex) |
| (7) | 1.23-1.54 | HI maps of high-velocity clouds and infrared maps of cirrus clouds |
| (8) | 1.34-1.40 | Molecular emission maps of clouds in the antigalactic center |
| (9) | 1.31-1.35 | Molecular emission maps (Ophiuchus, Perseus, and Orion clouds) |
| (10) | 1.50-1.53 | Molecular emission maps of clouds in the outer Galaxy |

Reference index: (1) Beech (1987); (2) Bazell and Desert (1988); (3) Dickman et al. (1990); (4) Falgarone et al. (1991); (5) Hetem and Lepine (1993); (6) Stutzki (1993); (7) Vogelaar and Wakker (1994); (8) Lee (2004); (9) Sánchez et al. (2007a); (10) Lee et al. (2008).

In general, observed values are spread over the range $1.1 \lesssim D_{per} \lesssim 1.5$. It is not clear, however, whether the different values seen in Figure 3 represent real variations of whether they are consequence of different observational data and/or analysis techniques. It is known that the obtained results may be affected by factors such as image resolution and/or signal-to-noise ratio (Dickman et al., 1990; Vogelaar and Wakker, 1994; Lee, 2004; Sánchez et al., 2005, 2007a). Note, for example, that for CO emission maps of the same region in the Taurus molecular complex Falgarone et al. (1991) obtained $D_{per} = 1.36$ whereas Stutzki (1993) found $D_{per} = 1.51$ on a different set of data. In order to get reliable clues about the ISM structure, it is important that any analysis technique is applied *systematically* on *homogeneous* data sets. Sánchez et al. (2007a) used several maps of different regions (Ophiuchus, Perseus, and Orion molecular clouds) in different emission lines and calculated $D_{per}$ by using an algorithm previously calibrated on simulated fractals (Sánchez et al., 2005). In this case the range of obtained values decreased notoriously to $1.31 \leq D_{per} \leq 1.35$. The general belief in this field (and we emphasize the term "belief") is that the fractal dimension of the projected boundaries of interstellar clouds is roughly a constant throughout the Galaxy, with $D_{per} \simeq 1.3 - 1.4$ (Bergin and Tafalla, 2007). This constancy in $D_{per}$ would be a natural consequence of a universal picture in which interstellar turbulence is driven by the same physical mechanisms everywhere (Elmegreen and Scalo, 2004).

But what is the value of the fractal dimension of interstellar clouds in the three-dimensional space, $D_f$? It has been traditionally assumed that $D_f = D_{per} + 1 \simeq 2.3 - 2.4$ (Beech, 1992). The fractal dimension of a very thin slice extracted from the three-dimensional distribution is $D_{slice} = D_f - 1$ (Falconer, 1990), but clouds are observed as projected images and a projection is a totally different operation. In fact, according to Falconer (1990) the expected dimension for projected fractals is $D_{projection} = min\{2, D_f\}$ so that from this relation we should obtain $D_f \simeq 1.3 - 1.4$ for the ISM. However, these equations cannot be used for estimating $D_f$ from $D_{per}$ because they refer to *internal* fractal dimensions, that is, to the way in which the object fills the space it occupies. The dimension of the *boundary* tells us how the projected contours (or the surface in 3D) fill the space. Both dimensions must be closely related to each other (at least for perfect monofractals) but as an object becomes more fractal $D_f$ decreases and $D_{per}$ increases, contrary to the above relations. Sánchez et al. (2005) used simulated fractal clouds to study the relationship between $D_{per}$ and $D_f$. Their results, summarized here in Figure 4, indicate that if the perimeter dimension is around $D_{per} \simeq 1.31 - 1.35$ (Sánchez et al., 2007a) then the corresponding 3D fractal dimension should be in the range $D_f \sim 2.6 - 2.8$. This dimension is higher than the value $D_f \sim 2.3$ that is usually

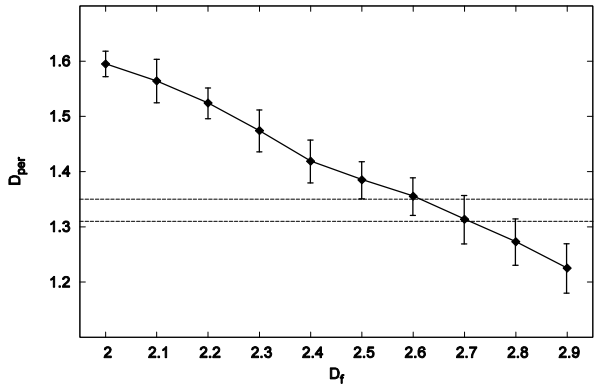

Figure 4: Perimeter dimension $D_{per}$ calculated on projected images for three-dimensional simulated clouds with fractal dimensions $D_f$. Horizontal dashed lines indicate the range of $D_{per}$ values estimated for different interstellar clouds in the Galaxy by Sánchez et al. (2007a).

assumed in the literature for interstellar clouds (Bergin and Tafalla, 2007).

The comparison of these results with models and simulations may give important clues on the physical process involved in the structure of the ISM. Numerical simulations of supersonic turbulence by Federrath et al. (2009) indicate that the resulting gas distributions may be very different depending on the way energy driving the turbulent motions is injected. Some physical mechanisms tend to produce what is known as compressive or irrotational flow (with no "twist") such as, for instance, spiral density shocks at galactic level. Other energy transfer mechanisms are mainly solenoidal or rotational, as the differential rotation in the Galactic disc. Federrath et al. (2009) found that compressive forcing produces much stronger density enhancements than solenoidal forcing. Interestingly, their simulations (for which the mach number was fixed at 5.5) yielded 3D fractal dimensions of $D_f \sim 2.3$ for the compressive mode and $D_f \sim 2.6$ for solenoidal forcing. Given the large number of physical mechanisms that can create turbulent motions in the ISM (Elmegreen and Scalo, 2004), it is likely that both compressive and solenoidal forcings coexist, but this kind of analysis is essential if we want to understand the origin of the ISM structure.

## 3. HIERARCHICAL AND SELF-SIMILAR STAR FORMATION

The distribution of stars and star-forming regions also exhibits a spatial hierarchy. Most stars are born grouped in star clusters and associations. But these clusters tend to be grouped together into larger "aggregates" of several clusters or even larger "star complexes" with spatial scales of the order of a kiloparsec (see, for example, Efremov, 1995; de la Fuente Marcos and de la Fuente Marcos, 2006, 2009; Elias et al., 2009; Elmegreen,

2010). Star clusters are not in the lower level of this hierarchy because many open clusters, mainly young embedded clusters, also exhibit smaller substructures in the distribution of their stars (Lada and Lada, 2003; Elmegreen, 2010). This fractal structure is presumably a direct consequence of the fact that stars are formed in a medium with an underlying fractal structure (previous Section). If this were the case, then it is reasonable to assume that the fractal dimension of the distribution of new-born stars should be nearly the same as that of the molecular clouds from which they are formed.

When dealing with a distribution of points (e.g., stars) in space, it is very useful to estimate the fractal dimension by using the so-called correlation integral $C(r)$, which is simply the probability of finding a point within a circle of radius $r$ centered on another point. For a fractal set it holds that $C(r) \sim r^{D_c}$, being $D_c$ the fractal dimension of the point distribution that, in this case, is called the correlation dimension. As mentioned before, for a homogeneous distribution of stars in the plane of the sky we expect $D_c = 2$, whereas if the stars are distributed obeying a fractal geometry with clumpy patterns then $D_c < 2$. The mean surface density of companions (MSDC) per star $\Sigma(\vartheta)$ as a function of angular separation $\vartheta$ is another widely used way to measure the degree of clustering of stars. For fractals $\Sigma(\vartheta) \sim \vartheta^\gamma$, and the exponent is related to the fractal dimension through $D_c = 2 + \gamma$.

MSDC technique has been used by various authors to study the clustering of protostars, pre-main sequence stars, or young stars in different star-forming regions. Most results seem to indicate that there are two different ranges of spatial scales, the regime of binary and multiple systems on smaller scales and a regime of fractal clustering on the largest scales. The idea prevalent among astronomers is that self-similar clustering above the binary regime is due to, or arises from, the fractal features of the parent clouds. For example, by using MSDC, Larson (1995) obtained $D_c = 1.38$ in the range $0.04 \lesssim r \lesssim 5$ pc in the Taurus-Auriga region. The spatial distribution of clumps in the star-forming region NGC 6334 reveals self-similar clustering with the same dimension $D_c = 1.38$ (Muñoz et al., 2007). However, such as in the case of gas distribution in GMCs, if one checks the references a wide variety of different values can be found. Nakajima et al. (1998) found significantly variations among different star-forming regions with $1.2 \lesssim D_c \lesssim 1.9$. There can be large differences even in the same regions if analyzed by different authors and data sets. For example, in the Taurus region both Larson (1995) and Simon (1997) analyzed young stars and their results are in perfect agreement with $D_c \simeq 1.4$, whereas Hartmann (2002) and Kraus and Hillenbrand (2008) both agree in $D_c \sim 1.0$ for the same region. Table 2 summarizes main results obtained by using the MSDC technique, where we can see the wide range of estimated $D_c$ values. Obviously, there can be different results depending on data sources, object selection criteria, and details of the specific calculation procedures. Additionally, it has been shown that if boundary and/or small data-set effects are not taken into account the final results can be seriously biased, given fractal dimension values smaller than the true ones (Sánchez et al., 2007b; see also Sánchez and Alfaro, 2008). In Figure 5 we have plotted $D_c$ values from Table 2 as a function of the number of data points $N_{dat}$. The observed behavior seems to be biased (at least in part) in the sense that $D_c$ decreases as $N_{dat}$ decreases (compare Figure 5 here with Figures 2 and 4 in Sánchez and Alfaro 2008). If this kind of effect is not corrected then any real variation in $D_c$ could be hidden or misunderstood.

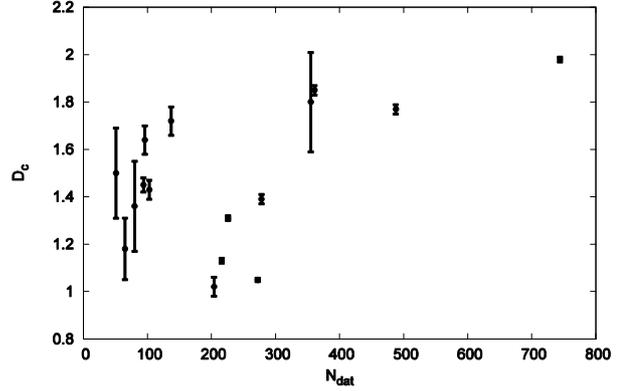

Figure 5: Fractal (correlation) dimension for the distribution of young stars and pre-main-sequence stars in different star-forming regions as a function of the number of data points (from Table 2).

Table 2: Summary of correlation dimensions for the distribution of stars in clusters (literature results that use the MSDC technique).

| Cluster | $N_{dat}$ | $D_c$ | Ref. |
|---|---|---|---|
| Taurus | 80 | 1.36± 0.19 | (1) |
| Taurus | 216 | 1.13±0.01 | (2) |
| Taurus | 204 | 1.02±0.04 | (3) |
| Taurus-Auriga | >121 | 1.38 | (4) |
| Taurus-Auriga | 272 | 1.049±0.007 | (5) |
| Orion A | 488 | 1.77±0.02 | (6) |
| Orion B | 226 | 1.31±0.01 | (6) |
| Orion OB | 361 | 1.85±0.02 | (6) |
| Trapezium | 355 | 1.80±0.21 | (1) |
| Trapezium | 744 | 1.98±0.01 | (7) |
| Ophiuchus | 51 | 1.50±0.19 | (1) |
| Ophiuchus | 96 | 1.64±0.06 | (6) |
| Chamaeleon | 94 | 1.45±0.03 | (6) |
| Chamaeleon I | 103 | 1.43±0.04 | (6) |
| Chamaeleon I | 137 | 1.72±0.06 | (2) |
| Vela | 278 | 1.39±0.02 | (6) |
| Lupus | 65 | 1.18±0.13 | (6) |

Reference index: (1) Simon (1997); (2) Gladwin et al. (1999); (3) Hartmann (2002); (4) Larson (1995); (5) Kraus and Hillenbrand (2008); (6) Nakajima et al. (1998); (7) Bate et al. (1998), their first data set.

## 4. EVOLUTIONARY EFFECTS

Even in the case of extremely young cluster, we always are seeing a snapshot of the cluster at a particular age resulting from certain initial conditions (ISM structure) and early dynamical evolution. As a cluster evolves, its initial distribution of stars may be erased, or at least modified.

When a star cluster is born, it contains an internal "thermal" energy (or total kinetic energy) that is given by the star velocity dispersion $v_{disp}$. This energy will tend to expand the cluster. This tendency is opposed by the gravitational potential of the total mass $M_{tot}$ that favors the collapse. If a cluster is in virial equilibrium between these two tendencies then $v_{disp}^2 \sim M_{tot}/R_c$, being $R_c$ the cluster radius. At the moment of formation $M_{tot}$ includes both mass in stars $M_{stars}$ and mass in remaining gas $M_{gas}$. However, just after the formation process stellar feedback through radiation, winds and outflows rapidly removes a large amount of gas and then $M_{tot}$ decreases abruptly. The subsequent evolution of the cluster will depend, among other things, on how much gas was removed, i.e. on the star formation efficiency $\epsilon = M_{stars}/M_{tot}$ (Gieles, 2010). If $\epsilon$ is below a threshold value the system likely will be gravitationally *unbound*. Simple theoretical considerations suggest that this limit is $\epsilon \leq 0.5$, although it is likely smaller than this (Gieles, 2010). In a gravitationally unbound cluster, the separation of the stars increases with age until the cluster dissolves into the field. In principle, the initial clumpy structure disappears after this process of "thermal" expansion, although some simulations suggest that it is possible to keep the initial substructure for a long time in unbound clusters (Goodwin and Whitworth, 2004). The opposite situation corresponds to clusters that after gas removal are gravitationally *bound*. These clusters have to evolve toward a new equilibrium state. Simulations show that this dynamical evolution can be a very complex process (e.g., Moeckel and Bate, 2010). It seems that the general trend is to evolve from the initially substructured distribution of stars toward centrally peaked distributions, that is, radial star density profiles. The evidence for this kind of evolution comes from both observations and from numerical simulations (Schmeja and Klessen, 2006; Schmeja et al., 2008a; Sánchez and Alfaro, 2009; Allison et al., 2009, 2010; Moeckel and Bate, 2010).

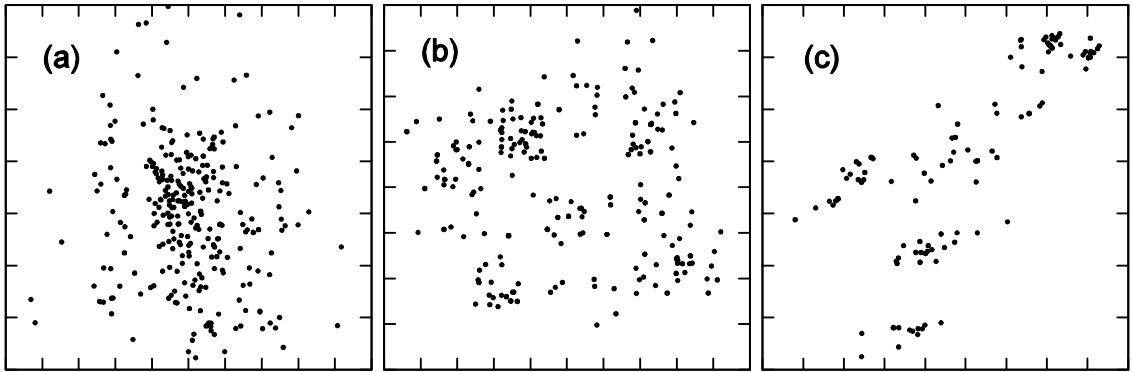

Figure 6: Positions relative to the cluster center for low-mass stars and brown dwarfs in the regions of the open clusters (a) IC 348 (data taken from Luhman et al., 2003), (b) IC 2391 (data from Barrado y Navascués et al., 2001) and (c) Taurus (Hartmann, 2002).

Figure 6 shows the distribution of low-mass stars in three example open clusters. The stars in the cluster IC 348 are concentrated toward the center following a nearly radial distribution. IC 2391 exhibits a more homogeneous distribution but with certain degree of clumpiness, i.e., several concentrations or "miniclusters" of stars. The distribution of stars in Taurus is clumpier than in IC 2391. The general idea is that these structure differences are due to evolutionary effects.

Roughly speaking, the time interval necessary to erase any initial structure will depend on the star velocity dispersion and the cluster size. The crossing time $T_{cross}$ is the time it takes a star to cross the cluster and it is given by $T_{cross} \sim R_c / v_{disp}$. The longer the propagation of information through the cluster (higher radius and/or smaller star velocities), longer time is required to erase the initial state. It should take at least several crossing times to reach an equilibrium state and/or to eliminate the original distribution (Goodwin and Whitworth, 2004), although some simulations indicate that the evolution from clumpy to radial distribution may occur on time scales as short as ~ 1 Myr (Allison et al., 2009, 2010). In order to address these questions, it is necessary to characterize the internal structure of young clusters and also to get some idea about the evolutionary stage (something related to the cluster age in crossing time units).

## 5. CHARACTERIZING THE DISTRIBUTION OF STARS

For radially concentrated clusters (e.g., IC 348 in Figure 6), star distribution is usually characterized by fitting the density profile (the number of stars per unit area versus the distance from the center) to some given predefined function. From the fitting procedure, we get parameters such as the central density of stars, the steepness of the density profile, and cluster radius. Obviously, this kind of analysis does not work in clumpy clusters (e.g., boxes b and c in Figure 6) because a smooth function cannot be well fitted to an irregular distribution. In this case, it is preferable to measure the degree of clustering through, for example, the fractal dimension $D_c$.

Cartwright and Whitworth (2004) proposed a different method to quantify the internal structure of star clusters. Their technique is becoming very widespread and useful for analyzing both observational and simulated data because, even being very simple and straightforward, it is able to quantify the structure and to distinguish between centrally concentrated and fractal-like distributions. The technique is based on the construction of the minimum spanning tree (MST). The MST is the set of straight lines (called branches or edges) connecting a given set of points without closed loops, such that the total edge length is minimum (see Figure 7). If we call $m$ the (normalized) mean length of branches in the MST and $s$ the (normalized) mean separation between stars, then an adimensional structure parameter can be defined as $Q = m/s$ (Cartwright and Whitworth, 2004; see also Schmeja and Klessen, 2006). For an homogeneous distribution of stars $Q \simeq 0.8$. If, instead, the stars are distributed in a clustered way then both $m$ and $s$ decrease because the separation among stars decreases, but interestingly the decreasing of $m$ and $s$ are different (either faster or lower) depending on the specific type of clustering. The behavior is such that $Q > 0.8$ for radial clustering whereas $Q < 0.8$ for fractal clustering (see the examples in Figure 7). Moreover, $Q$ increases as the steepness of the profile increases for radial clustering and $Q$ decreases as the fractal dimension decreases for fractal-type clustering (see Figure 5 in Cartwright and Whitworth, 2004; and Figure 7 in Sánchez and Alfaro, 2009). Thus, $Q$ is able to disentangle between radial and fractal clustering but it also measure the strength of clustering.

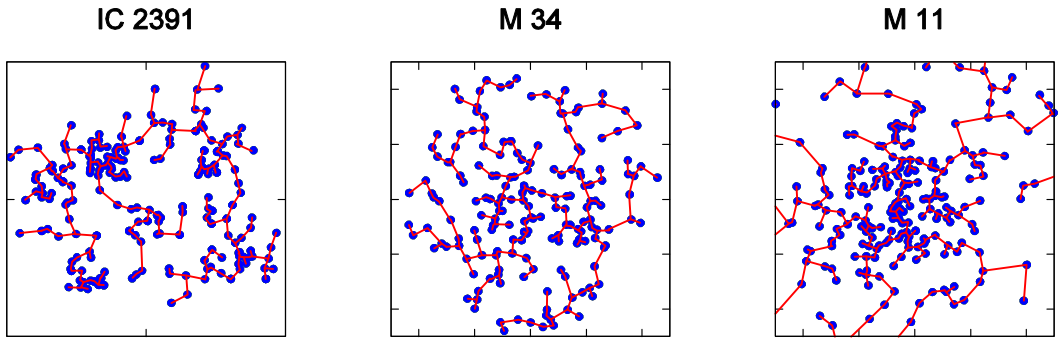

Figure 7: Minimun spanning trees for three open clusters, from which the parameter $Q$ can be calculated. Star positions are indicated with blue circles and red lines represent the trees. The value of $Q$ quantifies the spatial distribution of stars. For IC 2391 the stars are distributed following an irregular, fractal pattern with ($Q = 0.66 < 0.8$), for M 34 the stars are distributed roughly homogeneously ($Q = 0.8$), and for M 11 the stars follow a radial density profile ($Q = 1.02 > 0.8$).

Based on the previous discussion (Sections 3 and 4) and on the behavior of $Q$, we expect that the internal structure of a star cluster evolves with time from initial fractal clustering ($Q < 0.8$) to either homogeneous distribution ($Q \simeq 0.8$) if the cluster is dispersing its stars or centrally concentrated distribution ($Q > 0.8$) if it is a bound cluster. Cartwright and Whitworth (2004) were the first to calculate $Q$ in several star clusters. They obtained $Q = 0.47$ for stars in Taurus, a value consistent with its observed clumpy structure (panel c in Figure 6) and with its relatively young evolutionary stage (they estimated an age in crossing time units of $T/T_{cross} \simeq 0.1$). However, they also found some apparent contradictory results. IC 348, a slightly evolved cluster with $T/T_{cross} \simeq 1$ yielded $Q = 0.98$ according to its steep radial density profile (panel a in Figure 6). Instead, the highly evolved cluster IC 2391 ($T/T_{cross} \simeq 20$) still exhibits fractal clustering with $Q = 0.66$ (panel b in Figure 6). Schmeja et al. (2008a) applied this technique to embedded clusters in the Perseus, Serpens and Ophiuchus molecular clouds, and found that older Class 2/3 objects are more centrally condensed than the younger Class 0/1 protostars. Sánchez and Alfaro (2009) measured $Q$ in a sample of 16 open clusters spanning a wide range of ages. They found that there can exist clusters as old as ~ 100 Myr exhibiting fractal structure (e.g., NGC 1513 and NGC 1647). This means that either the initial clumpiness may last for a long time or other mechanisms may develop some kind of substructure starting from an initially more homogeneous state. From the analysis of a sample of embedded clusters and open clusters Schmeja et al. (2008b) suggested that clusters evolving from the embedded phase to the open cluster phase may regress to a more hierarchical configuration. These are actually real possibilities according to numerical simulations by Goodwin and Whitworth (2004), although some coherence in the initial velocity dispersion is required. Sánchez and Alfaro (2009) obtained a statistically significant correlation between $Q$ and $T/T_{cross}$ in a sample of open clusters. They calculated membership probabilities by applying, in a systematic and self-consistent way, a nonparametric method that does not make any assumption on the underlying star distribution. Their final sample consisted of 16 open clusters spanning a wide range of ages, from ~ 7.4 Myr to ~ 4.3 Gyr. Figure 8 shows the obtained correlation where the crossing times were calculated by assuming a constant velocity dispersion of 2 km s$^{-1}$. As we can see, the general trend is that young clusters (meaning that dynamically less evolved clusters) tend to distribute their stars following fractal patterns whereas older clusters tend to exhibit centrally concentrated structures. This result support the idea that stars in newly born clusters likely follow the fractal patterns of their parent molecular clouds, and that they eventually evolve towards more centrally concentrated structures. However, we know that this is only an overall trend. Some very young clusters may exhibit radial density gradients, as for instance σ Orionis for which $Q \simeq 0.9$ (Caballero, 2008).

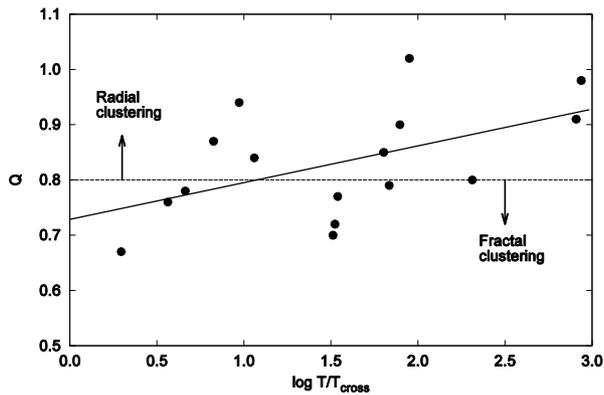

Figure 8: Structure parameter $Q$ as a function of the logarithm of age in crossing time units. The dashed line at $Q = 0.8$ separates radial from fractal clustering, and the solid line is the best linear fit.

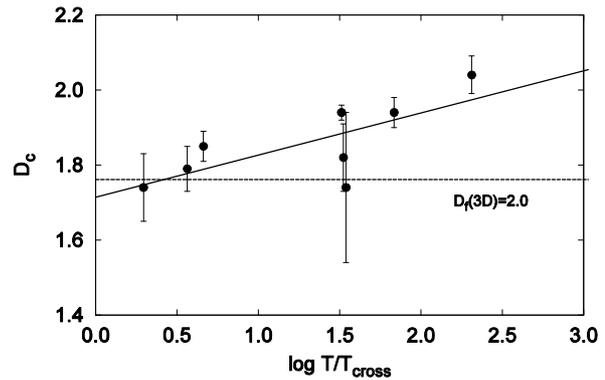

Figure 9: Correlation dimension as a function of age (in crossing-time units). The best linear fit is represented by a solid line. For reference, a horizontal dashed line indicates the value corresponding to three-dimensional distributions with fractal dimensions of $D_f = 2.0$.

Given the wide variety of physical processes involved in the origin and early evolution of star clusters, it is somewhat surprising that a correlation like that seen in Figure 8 can be observed. Very recent simulations by Allison et al. (2009, 2010) and Moeckel and Bate (2010) show that the transition from fractal clustering to central clustering may occur on very short timescales ($\lesssim 1$ Myr). Simulations by Maschberger et al. (2010) suggest a more complex variety of possibilities. They systematically analyzed two different simulations of turbulent fragmentation producing up to thousands of stars in clusters that can individually contain up to several hundred stars. Bound systems start fractal and evolve towards a centrally concentrated stage whereas unbound systems stay fractal in time. But this is the evolution for the *whole* systems. Star clusters in each system may evolve in totally different ways. In fact, the time evolution of the $Q$ parameter of clusters fluctuates dramatically depending on episodes of relaxation or merging (see Figure 8 in Maschberger et al., 2010, for viewing these behaviors). It is difficult to argue that, despite all this complex formation history (occurring in ~0.5 Myr), we should still observe some correlation between internal structure and age. In any case, a potentially important point is that now observers and simulators are both trying to use the same tools (in this case $Q$) for analyzing their results and this allows a better comparison and, therefore, understanding.

## 6. INITIAL FRACTAL STRUCTURE OF STAR CLUSTERS

For those clusters with fractal patterns, we can calculate the fractal dimension $D_c$ directly from the autocorrelation function (Section 3), which in general yields more accurate and precise results than inferring it from $Q$ (Sánchez and Alfaro, 2009). The result of doing this for open clusters with internal substructure (Sánchez and Alfaro, 2009) is shown in Figure 9. Again, a significant correlation is observed between $D_c$ and

$T/T_{cross}$. The degree of clumpiness is smaller for more evolved clusters. A horizontal dashed line in Figure 9 shows a reference $D_f$ value estimated from previous papers (see Figure 1 in Sánchez and Alfaro, 2008). To convert from two-dimensional $D_c$ values to three-dimensional $D_f$ values usually increases the associated uncertainties, so this reference value has to be taken with caution. However, from this figure is interesting to note that open clusters with the smallest correlation dimensions ($D_c = 1.74$) would have 3D fractal dimensions around $D_f \sim 2$. This value is considerably smaller than the average value estimated for Galactic molecular clouds (see Section 2), which is $D_f \simeq 2.6 - 2.8$.

This result creates an apparent problem, or at least a challenge to be addressed, because as mentioned before a group of stars born from the same cloud at almost the same place and time is expected to have a fractal dimension similar to that of the parent cloud. If the fractal dimension of the interstellar medium has a nearly *universal* value around 2.6–2.8, then how can some clusters exhibit such small fractal dimensions? This is still an open question. Several possibilities should be investigated in future studies. First, some simulations demonstrate that it is possible to increase the clumpiness (to decrease $D_f$) with time (Goodwin and Whitworth 2004), but it is necessary to understand the physical conditions under which this effect may occur. Second, maybe this difference is a consequence of a more clustered distribution of the densest gas from which stars form on the smallest spatial scales in the molecular cloud complexes, according to a multifractal scenario (Chappell and Scalo, 2001). If this were the case, different (smaller) fractal dimensions should be clearly measured in maps of very dense molecular regions. Third, perhaps the star formation process itself modifies in some (unknown) way the underlying geometry generating distributions of stars that can be very different from the distribution of gas in the star-forming cloud. Theoretical models or simulations are the way in which this idea can be explored. A fourth possibility is that the fractal dimension of the interstellar medium in the Galaxy does not have a universal value and therefore some clusters show smaller initial fractal dimensions because they formed in more clustered regions. Federrath et al. (2009) have shown that the fractal dimension of the ISM can be very different depending on the way in which turbulent energy is injected. Thus, $D_f$ could be very different from region to region in the Milky Way depending on the main physical processes driving the turbulence. Therefore, the possibility of a non-universal fractal dimension for the ISM should not, in principle, be ruled out. However, in this last case, overall correlations as those shown in Figures 8 and 9 should not, in principle, be observed. Obviously, it is also possible a combination of two or more of the above effects.

## 7. FINAL REMARKS

The internal structures of open clusters and associations give us important clues concerning their formation mechanism and dynamical evolution. There are both observational and numerical modeling evidences that open clusters evolve from an initial clumpy structure with several smaller subclusters toward a centrally condensed state. This simple picture has, however, several drawbacks. There can be very young clusters exhibiting radial patterns maybe reflecting the early effect of gravity on primordial gas. There can be also very evolved cluster showing fractal patterns that either have survived through time or have been generated subsequently by some (unknown) mechanism. Additionally, the corresponding 3D structure of some open clusters is much clumpier than the ISM gas distribution in the Milky Way; although in principle we would expect a very similar structure.

These points clearly require more investigation, but the problem is complex because it depends on: (1) the initial distribution of gas and dust in the parent cloud (the structure of GMCs), (2) the way and degree in which this information is transferred to the new-born stars (the star formation process), and (3) how, and how fast, this initial star distribution evolves (dynamical evolution of clusters). Each one of these factors is per se a research line topic with many physical processes involved. Moreover, it is necessary that any research in this area is done by measuring the cluster structures in an objective, quantitative, as well as systematic way. For this it is necessary to develop and use suitable analysis tools. The minimum spanning tree and fractal analysis seem to be very powerful techniques. It seems also to be particularly important to use data sets as homogeneous as possible, because the results may vary considerably for the same region depending on the data. This requirement will be at least partially satisfied by the upcoming ESA's mission GAIA which will make a complete census of stellar populations in the Galaxy down to 20th magnitude.

GAIA will work only at optical wavelengths and much of the galactic plane will be hidden by dust. It is worthwhile completing the studies with additional extensive observations at infrared wavelengths to detect and characterize embedded clusters.

## Acknowledgments

This work has made use of NASA's Astrophysics Data System. We acknowledge financial support from MICINN of Spain through grant AYA2007-64052 and from Consejería de Educación y Ciencia (Junta de Andalucía) through TIC-101 and TIC-4075. N.S. is supported by a post-doctoral JAE-Doc (CSIC) contract. E.J.A. acknowledges financial support from the Spanish MICINN under the Consolider-Ingenio 2010 Program grant CSD2006-00070: "First Science with the GTC".